\documentclass[twocolumn,11pt]{aastex631}
\usepackage{graphicx,amsmath}
\usepackage{color}
\usepackage{booktabs}
\usepackage{bm}

\usepackage{times}
\usepackage{float}
\usepackage{ulem}
\usepackage{multirow}
\usepackage{amsmath}

\newcommand{\adsurl}[1]{\href{#1}{ADS}}
\providecommand{\url}[1]{\href{#1}{#1}}

\newcommand{\be}{\begin{equation}}
\newcommand{\ee}{\end{equation}}
\newcommand{\bea}{\begin{eqnarray}}
\newcommand{\eea}{\end{eqnarray}}

\usepackage{color}
\usepackage{ifthen}
\newboolean{editorial}
\setboolean{editorial}{true}
\newcommand{\editorial}[2]{\ifthenelse{\boolean{editorial}}{\textcolor{red}{[\textsf{\textbf{{#1}}}: }\textcolor{blue}{\textsf{{#2}}}\textcolor{red}{]}}{}}

\shorttitle{Detection rate of sBBH mergers lensed by galaxy clusters}
\shortauthors{Chen, Xie, Lu, et al.}

\begin{document}
\title{Detection Rate of Galaxy Cluster Lensed Stellar Binary Black Hole Mergers by the Third-generation Gravitational Wave Detectors}
\correspondingauthor{Youjun Lu, Huanyuan Shan}
\author[0000-0001-7952-7945]{Zhiwei Chen}
\affiliation{National Astronomical Observatories, Chinese Academy of Sciences, 20A Datun Road, Beijing 100101, China}
\affiliation{School of Astronomy and Space Sciences, University of Chinese Academy of Sciences, 19A Yuquan Road, Beijing 100049, China}
\author{Yushan Xie}
\affiliation{Shanghai Astronomical Observatory ,Chinese Academy of Sciences, Nandan Road 80, Shanghai 200030, China}
\affiliation{School of Astronomy and Space Sciences, University of Chinese Academy of Sciences, 19A Yuquan Road, Beijing 100049, China}
\author[0000-0002-1310-4664]{Youjun Lu}\thanks{luyj@nao.cas.cn}
\affiliation{National Astronomical Observatories, Chinese Academy of Sciences, 20A Datun Road, Beijing 100101, China}
\affiliation{School of Astronomy and Space Sciences, University of Chinese Academy of Sciences, 19A Yuquan Road, Beijing 100049, China}
\author{Huanyuan Shan}\thanks{hyshan@shao.ac.cn}
\affiliation{Shanghai Astronomical Observatory ,Chinese Academy of Sciences, Nandan Road 80, Shanghai 200030, China}
\affiliation{Key Laboratory of Radio Astronomy and Technology, Chinese Academy of Sciences, A20 Datun Road, Beijing, 100101, China}
\affiliation{School of Astronomy and Space Sciences, University of Chinese Academy of Sciences, 19A Yuquan Road, Beijing 100049, China}
\author{Nan Li}
\affiliation{National Astronomical Observatories, Chinese Academy of Sciences, 20A Datun Road, Beijing 100101, China}
\affiliation{School of Astronomy and Space Sciences, University of Chinese Academy of Sciences, 19A Yuquan Road, Beijing 100049, China}
\author{Yuchao Luo}
\affiliation{National Astronomical Observatories, Chinese Academy of Sciences, 20A Datun Road, Beijing 100101, China}
\affiliation{School of Astronomy and Space Sciences, University of Chinese Academy of Sciences, 19A Yuquan Road, Beijing 100049, China}
\author[0000-0001-5174-0760]{Xiao Guo}
\affiliation{School of Fundamental Physics and Mathematical Sciences, Hangzhou Institute for Advanced Study, University of Chinese Academy of Sciences, Hangzhou 310024, China}

\begin{abstract}
Gravitational waves (GWs) from stellar binary black hole (sBBH) mergers can be strongly gravitational lensed by intervening galaxies/galaxy clusters. Only a few works investigated the cluster-lensed sBBH mergers by adopting oversimplified models, while galaxy-lensed ones were intensively studied. In this paper, we estimate the detection rate of cluuster-lensed sBBH mergers with the third-generation GW detectors and its dependence on the lens models. We adopt detailed modeling of galaxy cluster lenses by using the mock clusters in the Synthetic Sky Catalog for Dark Energy Science with LSST (CosmoDC2) and/or approximations of the pseudo-Jaffe profile or an eccentric Navarro-Frenk-White dark matter halo plus a bright central galaxy with singular isothermal sphere profile. Considering the formation of sBBH mergers dominates by the channel of evolution of massive binary stars (EMBS), we find that the detection rate of cluster-lensed sBBHs is $\sim5-84$\,yr$^{-1}$, depending on the adopted lens model and uncertainty in the merger rate density, and it is about $\sim{13_{-2.0}^{+28}}$\,yr$^{-1}$ if adopting relatively more realistic galaxy clusters with central main and member galaxies in the CosmoDC2 catalog, close to the estimated detection rate of sBBH mergers lensed by galaxies. In addition, we also consider the case that the production of sBBH mergers dominated by the dynamical interactions in dense stellar systems. We find that the detection rate of cluster-lensed sBBHs if from the dynamical channel is about $1.5$ times larger than that from the EMBS channel and the redshift distribution of former peaking at a higher redshift ($\sim3$) compared with that from latter ($\sim2$).   
\end{abstract}

\section{Introduction}

More than 100 stellar binary black hole (sBBH) mergers have been detected since the first detection of the gravitational wave (GW; sBBH merger event GW150914) by Laser Interferometer GW Observatories (LIGO) and Virgo \citep{2016PhRvL.116m1102A, 2019PhRvX...9c1040A, 2020arXiv201014527A, 2021arXiv211103606T, 2021arXiv211103634T}, which have already provided abundant information on their mass and spin distributions, and thus helped to study the evolution of massive binary stars (EMBS) and the dynamical interactions of compact stars in dense stellar systems. With increasing sensitivities, the planned third-generation GW detectors, such as the Einstein telescope (ET; \citealt{Hild_2011}) and the Cosmic Explorer (CE; \citealt[][]{2019BAAS...51g..35R}), are anticipated to detect at least $10^4-10^5$ sBBH merger events per year with redshift up to $z_{\rm s}\sim 10$. Among these events, a small fraction ($\sim10^{-3}-10^{-4}$) are anticipated to be gravitational lensed by intervening objects and systems, including primordial black holes, mini-dark matter halos, galaxies, and galaxy clusters. Detection of such lensed GW events is one of the main objectives for GW observations. 
 
Gravitational lensing of sBBH merger events has been studied intensively in the literature in both the geometrical optics and diffraction wave optics regimes {\citep[e.g.,][]{1996PhRvL..77.2875W, 1998PhRvL..80.1138N, 2003ApJ...595.1039T,2018PhRvD..98j4029D,2019A&A...627A.130D,Guo:2022dre, 2023PhRvD.107d3029C, 2023arXiv230404800L}}, and it was suggested that these lensed events could provide important information on the formation and evolution of sBBHs, as well as to distinguish different dark matter models and constrain the expansion history of the universe  \citep[e..g.,][]{2017NatCo...8.1148L, 2020MNRAS.498.3395H, 2022ApJ...940...17C,Guo:2022dre}. For the astrophysical application of lensed sBBH GW events, one may need first to answer whether it can be detected or not, therefore, it is of great significance to estimate the detection rate of strongly lensed sBBH mergers for (upcoming) GW detectors.

Most of the previous works focus on predicting the detection rate of sBBH mergers lensed by intervening galaxies with multiple images in the geometric optics regime \citep[e.g.,][]{ 2014JCAP...10..080B, Piorkowska:2013eww, 2015JCAP...12..006D, 2018MNRAS.476.2220L, 2019ApJ...874..139Y, PhysRevD.103.104055, 2021MNRAS.501.2451M, 2021ApJ...921..154W, 2022MNRAS.509.3772Y, 2023ApJ...953...36C}, but rarely discuss the case of galaxy cluster lensing systems \citep[c.f.,][]{2018MNRAS.475.3823S, 2018IAUS..338...98S, 2023MNRAS.520..702S}. The detection rate of these cluster lensed sBBH mergers could be significant and may even be comparable to that lensed by intervening galaxies due to the large cross-section of the galaxy clusters. \citet{2023MNRAS.520..702S} predicted the total detection rate of lensed GWs for LIGO and Virgo, including both the galaxy and galaxy cluster lens cases. They adopted a universal singular isothermal sphere (SIS) model to calculate the lensing cross-section for both galaxy and cluster lenses, without considering the differences between them. This assumption may be oversimplified since galaxy clusters and galaxies may have different mass density distributions, with the former having significant contributions from or even dominated by dark matter in their central regions and the latter dominated by stellar matters in their central regions.

In this paper, we adopt relatively more realistic models to estimate the detection rate of sBBH mergers lensed by galaxy clusters via the third-generation GW detectors. We model the mass distributions of the cluster lenses obtained from the mock cluster in COSMODC2 and/or from semi-analytical approximations by the pseudo-Jaffe profile or an eccentric NFW profile of a dark matter halo plus a central BCG with SIS profile. We then estimate the detection rates for different models and different sBBH formation channels. This paper is organized as follows. In Section~\ref{sec:method}, we briefly introduce the method for estimating the detection rate of cluster-lensed sBBH mergers. In Section~\ref{sec:results}, we present our main results. Discussions and conclusions are given in Section~ \ref{sec:con}. Throughout the paper, we adopt the cosmological parameters as $(h_0,\Omega_{\rm m},\Omega_\Lambda)=(0.68,0.31,0.69)$ \citep{Aghanim2020}.

\section{Methodology}
\label{sec:method}

The galaxy cluster-lensing of sBBH mergers is anticipated to occur mainly at redshift $z\lesssim 5$ as the sBBH merger rate density declines rapidly beyond $z_{\rm s}\sim 5$ \citep[e.g.,][]{2015ApJ...806..263D,2016Natur.534..512B,2020ApJ...891..141G,2021ApJ...912L..23R,2023MNRAS.523.5719R}. Therefore, one may estimate the detection rate  $\dot{N}^{\ell}$ by
\begin{equation}
\begin{aligned}
\dot{N}^{\ell}=\int dz_{\rm s} &\int dq\int dm_1 \frac{d^3\dot{N}_{\bullet\bullet}(z_{\rm s})}{dz_{\rm s}dm_1dq}\\
&\times P(\sqrt{\mu^{\ell}}\rho>\rho_0 \mid m_1,q,z_{\rm s})
\tau^{\ell}(z_{\rm s}).
\end{aligned}
\end{equation}
Here $d^3\dot{N}_{\bullet\bullet}(z_{\rm s})/dz_{\rm s} dm_1 dq$ represents the number of sBBH mergers per unit time in the redshift range of $z_{\rm s} \rightarrow z_{\rm s}+dz_{\rm s}$ and the sBBH primary mass range of $m_1\rightarrow  m_1+dm_1$ and mass ratio range of $q\rightarrow q+dq$, $\tau^{\ell}({z_{\rm s}})$ is the optical depth for the galaxy-cluster lensing of a event at $z_{\rm s}$. The function $P(\sqrt{\mu^{\ell}}\rho>\rho_0 \mid m_1,q,z_{\rm s})$ describes the fraction of lensed events with given source parameters of ($m_1,q,z_{\rm s}$) that can be detected considering the signal-to-noise (S/N) enhancement due to the magnification $\mu^{\ell}$ of the signals, $\rho$ and $\rho_0$ are the expected S/N if the source is not lensed and the detection threshold for GW signals. Note here that the value of $\rho_0$ of sBBH mergers is directly related with their mass spectrum, thus dependent on their formation channel. In principle, the total masses of sBBH mergers produced via the EMBS channel may be relatively smaller than that produced by the dynamical channel. 
Thus, the magnification bias $P(\sqrt{\mu^{\ell}}\rho>\rho_{0}\mid m_1,q,z_{\rm s})$ may be different for the lensed sBBH merger events produced by different formation channels. However, the third generation GW detectors have extremely high sensitivity, and they can detect almost all the sBBH mergers with redshift $z_{\rm s}<17$ \citep[e.g.,][]{2011CQGra..28i4013H,2019BAAS...51g..35R}.  {For example, about $\sim 96.5\%$ of the sBBH mergers formed via EMBS channel can be detected, if assuming $\rho_0=8$.} For this reason, we set $P(\sqrt{\mu^{\ell}}\rho>\rho_{0}\mid m_1,q,z_{\rm s})=1$ for the detection of strongly lensed sBBH mergers by the third generation GW detectors, though the actual S/N of any source depends on both the source and lens properties. The detection rate $\dot{N}^{\ell}$ can be then estimated as
\begin{equation}
\dot{N}^{\ell}=\int dz_{\rm s}\int dm_1\int dq\frac{d^3\dot{N}_{\bullet\bullet}(z_{\rm s})}{dz_{\rm s}dm_1dq}\tau^{\ell}(z_{\rm s}), 
\label{eq:Nl}
\end{equation}
without detailed consideration of the S/N values of the GW signals.

\subsection{Merger rate density}

The intrinsic number density distribution of the sBBH merger GW events $d^3\dot{N}_{\bullet\bullet}(z_{\rm s}) /dz_{\rm s}dm_1dq$ is directly related to $R_{\bullet\bullet} (z,m_1,q)$, representing the merger rate density of sBBH mergers at redshift $z$ in the primary mass range ($m_1$, $m_1+ dm_1$) and the mass ratio range ($q$, $q+dq$), i.e.,
\begin{equation}
\frac{d^3\dot{N}_{\bullet\bullet}}{d{m_1}dq dz}=\frac{{R_{\bullet\bullet}}(z,{m_1},q)}{1+z} \frac{dV_{\rm c}(z)}{dz},  
\label{eq:source}
\end{equation}
where the factor $1/(1+z)$ accounts for the time dilation, $V_{\rm c}$ is the volume of the universe. 

Two astrophysical channels may dominate the formation of cosmic sBBH mergers. One is the EMBS channel, with which sBBHs were formed as the end products of the evolution of close massive binary stars, then lose orbital energy by GW and finally merge. The other is the dynamical channel, with which the compact stars and binaries in dense stellar systems gravitational interact with each other, which leads to the formation of close sBBHs and subsequent mergers due to GW decay. The sBBH merger rate density from these two different channels can be estimated as follows (more detailed descriptions of the estimates in \citet{2017cao} and \citet{2021MNRAS.500.1421Z}).

\begin{itemize}

\item {\bf EMBS channel:} the merger rate density can be estimated as
\begin{eqnarray}
\boldsymbol{R^{B}_{\bullet\bullet }}\left(z_{\rm s}, m_1,q\right) &= &\int d \tau_{\mathrm{d}}  f_{\rm eff}  R_{\text {birth }}(m_{1}, {z^{\prime}}) \nonumber \\
& & \times P_{\tau_{\rm d}}\left(\tau_{\mathrm{d}}\right) P_{q}(q), 
\label{eq:Ptaud}
\end{eqnarray}
and 
\begin{eqnarray}
R_{\mathrm{birth}}\left(m_{1}, {z^{\prime}}\right) & = &\iint d m_{\ast} d Z \dot{\psi}\left(Z ; {z^{\prime}}\right) \phi\left(m_{\ast}\right) \nonumber \\ 
& & \times  \delta\left(m_{\ast}-g^{-1}\left(m_{1}, Z\right)\right), 
\end{eqnarray}
where $f_{\rm eff}$ denotes the formation efficiency of sBBHs, which can be calibrated by the local sBBH merger density given by LIGO/Virgo observations, $\dot{\psi}(Z; {z^{\prime}})$ is the cosmic star formation rate density (SFR) with metallicity $Z$ at formation redshift $z^{\prime}$, which is assumed to be able to separate the dependencies of metallicty and redshift into two independent functions. More detailed description can be seen in  \citet{2016Natur.534..512B}. The function of $\phi(m_{\ast})$ is the initial mass function (IMF), and $m_1=g(m_{\ast,Z})$ is the relationship between the initial zero-age main sequence star mass and the remnant mass given by \citet{2015MNRAS.451.4086S}. The function $P_{\tau_{\rm d}}$ describes the distribution of the time delay between the merger of sBBHs and the formation time of its progenitor binary stars. We set $P_{\tau_{\rm d}} \propto \tau^{-1}_{\rm d}$ with the minimum and maximum values of $\tau_{\rm d}$ as $50$\,Myr and the Hubble time, respectively. The distribution of mass ratio $q$ is assumed to $\propto q$ in the range from $0.5$ to $1$ \citep[see][]{2016Natur.534..512B}. 

\item {\bf Dynamical channel:} if sBBH mergers are produced by the dynamical channel, can be estimated by associating the formation of sBBHs to the formation and evolution of globular clusters \citep{2021MNRAS.500.1421Z}, i.e.,
\begin{equation}
\begin{aligned}
& {R^{ D}_{\bullet\bullet}}(z_{\rm s},m_1,q)= \left.\iiint \frac{d\dot{M}_{\mathrm{G}}}{d \log M_{\mathrm{H}}}\right|_{z(\tau)} \frac{1}{\left\langle M_{\mathrm{G}}\right\rangle} P\left(M_{\mathrm{G}}\right) \\
& \times R\left(r_{\mathrm{v}}, M_{\mathrm{G}}, \tau-t(z_{\rm s})\right) P(m_1)P(q) d M_{\mathrm{H}} d M_{\mathrm{G}} d \tau,
\end{aligned}
\end{equation}
where $\frac{d\dot{M}_{\mathrm{G}}}{d \log M_{\mathrm{H}}}$ is the comoving SFR in globular clusters per galaxies of a given halo mass $M_{\mathrm{H}}$ at given redshift $z(\tau)$ (or a given formation time $\tau$ ). The function $P(M_{\rm G})$ is the cluster initial mass function, ${\left\langle M_{\mathrm{G}}\right\rangle}$ is the mean initial mass of a globular cluster, while $R(r_{\rm v}, M_{\rm G}, t)$ is the merger rate of sBBHs in a globular cluster with initial virial radius $r_{\rm v}$ and mass $M_{\rm G}$ at time $t(z_{\rm s})$. In this paper, we assume the primary mass distribution of sBBH mergers $P(m_1)$ produced by the dynamical interactions in globular cluster follows the results given in  \citet{2018ApJ...866L...5R}. {As for the mass ratio distribution $P(q)$, the results obtained by numerical simulations \citep{2016PhRvD..93h4029R,2018ApJ...866L...5R} are quite similar with that formed via the EMBS channel, to meet the observational results estimated by LVK, which is $P(q)\propto q^{\beta_{q}}$ and $\beta_{q}=1.1_{-1.3}^{+1.7}$ \citep{2021arXiv211103634T}. Therefore, we assume it to be the same with that of the EMBS channel. Noticed that different choice of $\beta_q$ will not affect our results much, since most of the sBBH mergers can possess SNR larger than $8$. For example, if choose a uniform mass ratio, i.e., $\beta_q=0$, the detection fraction of total sBBH mergers will be $\sim 94.7\%$.  }

Additionally, the specific form for $\frac{\dot{M}_{\mathrm{G}}}{d \log M_{\mathrm{H}}}$ and $R(r_{\rm v}, M_{\rm G})$ are set to be the explicit function given in \citet{2018ApJ...866L...5R}. Among these mock clusters there are  $50\%$ form with $r_{\mathrm{v}} = 1$\,pc and the rest $50\%$ form with $r_{\mathrm{v}} = 2$\,pc.  Then the total rate is the summation of those from the two $v_{\rm v}$ cases.

\end{itemize}

\begin{figure}
\centering
\includegraphics[width=1.0\columnwidth]{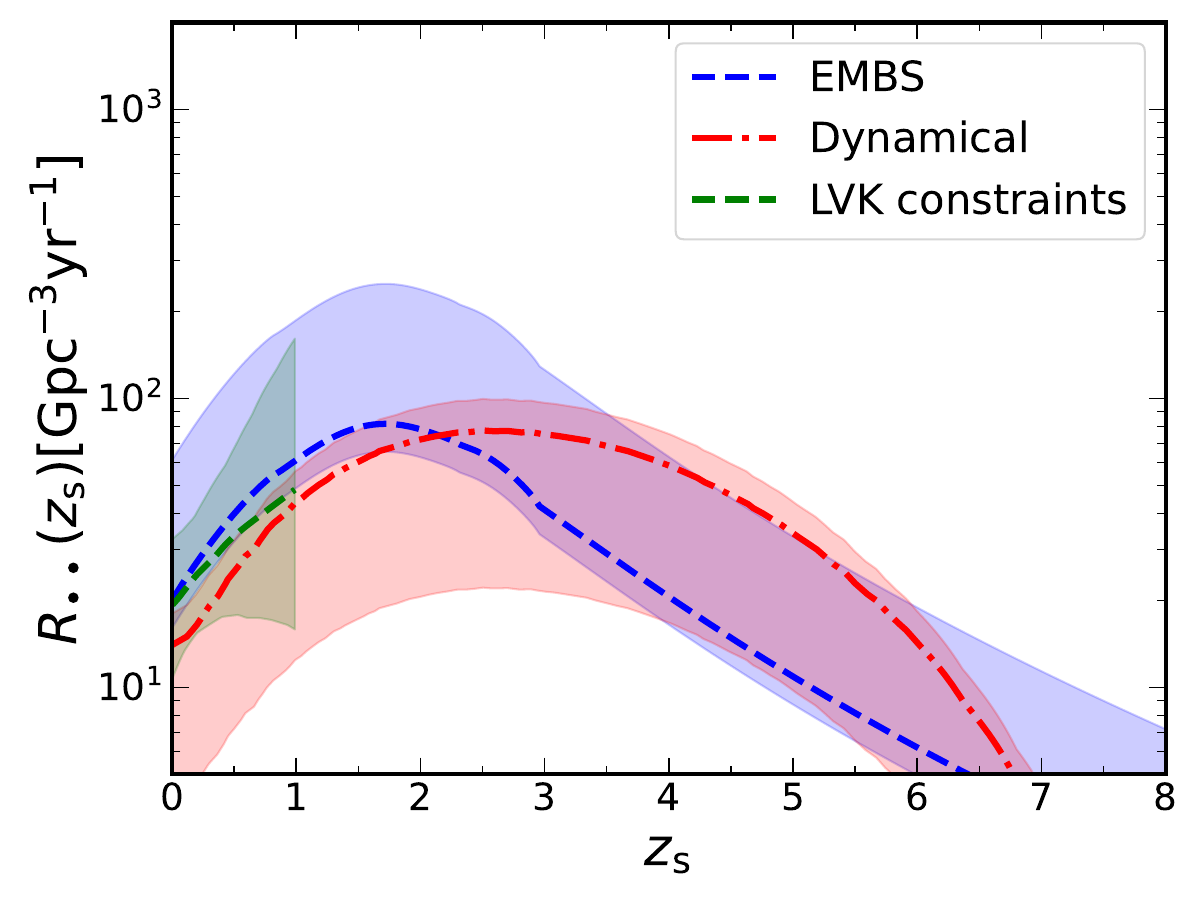}
\caption{The predicted redshift evolution of the merger rate density of sBBHs. The blue dashed and the red dotted-dashed lines show the results obtained from the EMBH and dynamical channels, respectively, with the calibration by the local sBBH merger rate density obtained from LIGO-Virgo observations, and the corresponding shaded regions show the errors induced by the uncertainties of local merger rate densities  ($90\%$ confidence interval) \citep[e.g., ][]{2018ApJ...866L...5R, 2021arXiv211103606T, 2021arXiv211103634T}. {Green shaded region shows the constraints on the merger rate density evolution obtained by the LVK observations assuming the peak+power law model \citep{2021arXiv211103634T}. 
}}
\label{fig:rbb}
\end{figure}

Figure~\ref{fig:rbb} shows the results of the sBBH merger rate density evolution for both the EMBH (blue dashed line) and dynamical (red dotted-dashed line) channels obtained from the above settings. The merger rate density evolution obtained for both channels is calibrated by the median local merger rate density estimated by LIGO-Virgo observations, i.e.,  $\sim 14\rm Gpc^{-3}yr^{-1}$ \citep[e.g., ][]{2018ApJ...866L...5R} and $\sim 19\rm Gpc^{-3}yr^{-1}$ \citep[e.g., ][]{2021arXiv211103606T, 2021arXiv211103634T}, respectively. The corresponding shade regions represent the errors induced by the uncertainties in the constraint of the local merger rate density, $ 4-18\rm Gpc^{-3}yr^{-1}$ and $ 16-61\rm Gpc^{-3}$\,yr$^{-1}$  ($90\%$ confidence interval) for the dynamical and EMBS channels, respectively. The merger rate density from the EMBS channel peaks at $z_{\rm s}\sim 1.5$ and has a value of $\sim 80 \rm Gpc^{-3}yr^{-1}$, while that from the dynamical and channel peaks at a higher redshift, i.e.,  $z_{\rm s}\sim 2.5$ and has a lower value of $\sim 75 \rm Gpc^{-3}yr^{-1}$. { We also show the merger rate density evolution obtained from the LVK observations (green shaded region) by assuming a peak+power law model with uncertainties due to the limited number of events and the lack of GW detection at higher redshift \citep[see][]{2021arXiv211103606T, 2021arXiv211103634T}. Our estimation on the merger rate desnity lies within such a large region.
}

\subsection{Optical depth: $\tau^{\ell}({z_{\rm s}})$}

The cosmic optical depth $\tau^{\ell}({z_{\rm s}})$  for the strong lensing caused by galaxy clusters can be estimated as
\begin{equation}
\tau^{\ell}({z_{\rm s}})=\frac{1}{4\pi}\int_{0}^{z_{\rm s}} dz_{\ell}\int dM_{\ell} S_{\rm cr}(z_\ell,z_{\rm s},M_{\ell})\frac{dn_{\rm H}}{dM_{\ell}}\frac{dV_{\rm c}}{dz_{\ell}},
\label{eq:tau}
\end{equation}  
where $z_{\ell}$ and $M_\ell$ are the redshift and the mass of the galaxy cluster, $dn_{\rm H}/dM_{\ell}$ denotes the halo mass function (HMF), and $S_{\rm cr}(z_\ell,z_{\rm s},M_{\ell})$ denotes the lensing cross-section of the cluster. Note that in the above expression for $\tau^{\ell}({z_{\rm s}})$, it is implicitly assumed that the lensing cross-section of the galaxy cluster is mainly determined by the mass distribution in its dark matter halo and the associated central galaxy, which is reasonable since the member galaxies only offer small perturbation to the caustic shape and/or add extra small caustics associated with each of them.

\subsubsection{Cluster lens model}

We estimate $S_{\rm cr}$ below by using two different approaches. One is a semi-analytic model by assuming a universal lens density profile (either the pseudo-Jaffe (PSJ) density profile \citep{1983MNRAS.202..995J} or an eccentric NFW density profile  \citep{1996ApJ...462..563N} for the dark matter halo plus a SIS density profile \citep{1994A&A...284..285K} for the central BCG). For this approach, we estimate the cross-section distribution by the dimensionless Einstein radius $y_{\rm cr}$ in the source plane as
\begin{equation}
S_{\rm cr}=\pi \left(\frac{\xi_{0}y_{\rm cr}}{D_{\ell}}\right)^2,
\label{eq:cs}
\end{equation}
where $\xi_{0}$ is the characteristic length in the lens plane and $D_{\ell}$ is the angular diameter distance of the galaxy cluster. 

The other is a model based on the numerical simulations of cluster lenses from the CosmoDC2 catalog, extending to larger mass and redshift ranges by using interpolation and extrapolation, which presumably gives a more realistic estimation on $\dot{N}^{\ell}$. In this case, the lensing cross-section of the mock clusters is calculated by ray tracing numerically \citep[e.g.,][]{2009MNRAS.395.1319J, 2015ApJ...813..102B, 2024MNRAS.531.1179X}, and we introduce a simple fitting form according to the numerical results for these mock clusters and then use it to obtain the cross-section for clusters by interpolation and extrapolation. 

\begin{itemize}
\item {\bf Pseudo-Jaffe profile: } A cluster lens, composed of the main dark matter halo and a central BCG, may be treated as a whole and assumed to follow the PSJ profile as 
\begin{equation}
\rho_{\rm PSJ}(r)=\frac{\rho_{\rm PSJ} \xi^4_{0,\rm PSJ}}{(r^2+s^2)(r^2+a^2)},
\end{equation}
where $\rho_{\rm PSJ}$ is the characteristic density, $s$ and $a$ are the core and transition radius respectively. By analogizing to the SIS model, the characteristic scale radius $\xi_{0,\rm PSJ}$ can be calculated by
\begin{equation}
\xi^2_{0,\rm PSJ}=\frac{4(x_a+x_s)GM_{\ell}D_{\rm eff}}{x_a^2c^2},
\label{eq:psj}
\end{equation}
where $(x,x_s,x_a)$=$(r,s,a)\xi_{0,\rm PSJ}^{-1}$, $D_{\rm eff}=D_{\ell}D_{\rm s}/D_{{\ell}s}$ is the effective angular diameter distance of the lens system, $D_{\ell}$, $D_{\rm s}$ and $D_{{\ell}s}$ are the angular diameter distance of the lens, source, and that from lens to source.

As discussed in \citet{2024MNRAS.531.1179X}, real clusters observed in the Hubble-frontier field can be modeled by the pseudo-isothermal elliptical mass distribution (PIEMD) \citep{2015MNRAS.452.1437J,lotz17,annu17,2018arXiv180600698Q}, of which the mass density profile is the same as the PSJ profile with eccentricity. \citet{2024MNRAS.531.1179X} found that the value of $x_s$ and $x_a$ can vary from $0.05-0.3$ and $1-3$. Note that the different choices of $x_s$ and $x_a$ may lead to the change $y_{\rm crit}$ by a factor of $2-5$. For  demonstration purpose, we simply assume i.e., $(x_s,x_a)=(0.2,2)$ and find that $y_{\rm cr}$ is $\sim 0.17$. 

 More detailed description of the PSJ model can be seen in the Appendix. 

\item \textbf{eNFW+SIS Profile}: the mass density of the NFW profile is given by 
\begin{equation}
\rho_{\rm NFW}(r)=\frac{\rho_{\rm NFW}}{\frac{r}{r_{\rm s}}\left(1+\frac{r}{r_{\rm s}}\right)^2},
\label{eq:nfw}
\end{equation}
where $\rho_{\rm NFW}$ is the characteristic density, $r_{\rm s}=r_{\rm vir}/c_{\rm v}$ is the characteristic length, which can be estimated according to the mass $M_\ell$, concentration $c_{\rm v}$, and virial radius $r_{\rm vir}$ of the dark matter halo. Note that the estimate of the optical depth is strongly dependent on the adopted relationship between $c_{\rm v}$ and  $M_\ell$. A simple intuition is that the higher concentration $c_{\rm v}$ with the same $M_\ell$, the larger the cross-section due to relatively stronger deflection of the GW. In this paper, we adopt two different fitting formulae in the literature, i.e., the larger $c^{\rm o}_{\rm v}$ \citep{2001ApJ...559..572O} and the smaller $c^{\rm w}_{\rm v}$ \citep{2020Natur.585...39W} relation given by
\begin{equation}
c^{\rm o}_{\rm v}(M_{\ell},z_\ell)=\frac{8}{1+z_{\ell}}\left(\frac{M_\ell}{10^{14}h_0^{-1}M_{\odot}}\right)^{-0.13},
\end{equation}
and 
\begin{equation}
\label{eq:c_vWang}
c^{\rm w}_{\rm v}=\frac{1}{1+z_{\rm s}}\exp \left[c_{6} \left(\frac{M_{\mathrm{fs}}}{M_{\ell}}\right)^{\frac{1}{3}}\right] \times \sum_{i=0}^{5} c_{\mathrm{i}}\left[\ln \frac{M_{\ell}}{h_0^{-1} M_{\odot}}\right]^{i},
\end{equation}
as the upper and lower limits, respectively. Here $h_0$ is the dimensionless Hubble constant, $c_i=\left[27.112,-0.381,-1.853 \times 10^{-3},-4.141 \times\right.$ $\left.10^{-4},-4.334 \times 10^{-6}, 3.208 \times 10^{-7},-0.529\right]$ for $i \in\{0, \ldots, 6\}$ and the free-streaming mass scale $M_{\mathrm{fs}}=7.3 \times 10^{-6} M_{\odot}$. Moreover, we assume the NFW profile to be eccentric with axis ratio   $\lambda\sim 0.3$ by replacing the radius $r$ defined in equation~\ref{eq:nfw} to $\xi=x^2+y^2/\lambda^2$  in Cartesian coordinate \citep[details see][]{1990A&A...231...19S}, which is the mean value of the mock samples from CosmoDC2, which will be introduced later. 

The central BCG is assumed to follow the SIS density profile as 
\begin{equation}
\rho_{\rm SIS}(r)=\frac{\sigma_{\rm v}^2}{2\pi G r^2},
\end{equation}
where $\sigma_{\rm v}$ is the velocity dispersion of the BCG. We link the BCG to the main dark matter halo by adopting the empirical relationship between their masses given by \citep{2019A&A...631A.175E}
\begin{equation}
\log ({{M_{\ast,\rm BCG}}/{M_{\odot}} })=t_1\log (M_{\ell}/M_{\odot})+t_2, 
\end{equation}

where $(t_1,t_2)$ are $(0.41,5.59)$ for $z_{\ell} \in [0.1,0.3]$ and $(0.31,7.00)$ for $z_{\ell} \in (0.3,0.65]$. Due to observation difficulties, the data points are not sufficient at the higher redshift outside of these two intervals to obtain such a relationship between $M_{\ast,\rm BCG}$ and $M_\ell$. {However, we may get insights from the results of cosmological simulations. As discussed in \citet{2022ApJ...938....3S}, they derived a similar relation from the IllustisTNG Snapshots and found  that this relation may not evolve much up to $z\sim 2$. 
Therefore, we simply assume the obtained relationships at $z_{\ell} \in [0.1,0.3]$ and $(0.31,7.00)$ can be extended to lower and higher redshift. This simple extension may introduce uncertainties of $1-3$ times to the detection rate estimation.
} 
Then the velocity dispersion $\sigma_{\rm v}$ of the BCG is estimated by the following relationship,

\begin{equation}
\log (\sigma_{\rm v})=0.286\log (M_{\ast, \rm BCG}/M_{\odot})-0.86,
\end{equation}
which is fitted by the observation of about $50000$ early-type galaxies from SDSS \citep{2009MNRAS.394.1978H}. In this way, we assume the characteristic scale radius $\xi_{0,\rm eNFW+SIS}$ to be the Einstein radius $R_{\rm E}$ of the BCG, 
\begin{equation}
R_{\rm E}=4\pi\left(\frac{\sigma_{\rm v}}{c}\right)^2\frac{D_{{\ell}s}}{D_{\rm s}}.
\end{equation}
Note that due to the complex density distribution of the lens system, i.e., the eNFW+SIS profile, it is difficult to solve the dependence of $y_{\rm cr}$ on $M_{\ell}$ and $z_\ell$ analytically. Though very time-consuming, we pixelize the source plane for each system and solve the ray-tracing equation pixel by pixel numerically with the usage of \textbf{LENSTRONOMY} \citep{2015ApJ...813..102B}. Then the cross-section $S_{\rm cr}$ for the eNFW+SIS profile is obtained by the Monte-Carlo method. 

\item {\bf CosmoDC2 clusters:} we also incorporate the mock galaxy clusters from the CosmoDC2 catalog \citep{korytov19} as the lensing models to account for a relatively more realistic case. The CosmoDC2 catalog is originally created for the LSST Dark Energy Science Collaboration \citep{lsst12}, playing an important role in various scientific goals. The catalog is constructed based on a large volume cosmological N-body simulation, the `Outer Rim', while the galaxies within are populated with empirical models using the Universe Machine \citep{behroozi19} and semi-analytic models using the Galacticus \citep{benson12} code. 

\begin{figure}
\centering
\includegraphics[width=1.0\columnwidth]{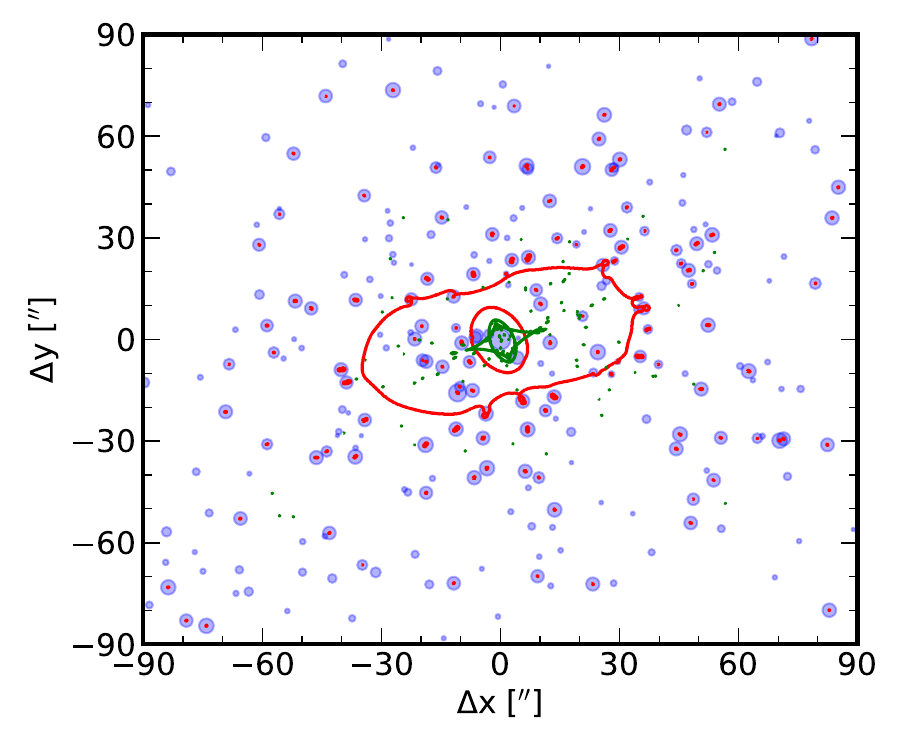}
\caption{A cluster example extracted from the CosmoDC2 catalog. Within the galaxy cluster, the positions of member galaxies are indicated as the blue filled circles, where the size of circles is proportional to the galaxy's absolute luminosity. The critical curves (caustics) for a source at $z_{\rm s} = 2$ are over-plotted as the red (green) lines.}
\label{fig:clus}
\end{figure}

We select well-modeled 3615 galaxy clusters with masses $ M_{\ell} \geq 10^{14} M_{\odot}$ in the redshift range $z\in (0, 1)$ from this catalog as the strong lenses. 
{
Note that clusters at higher redshifts are not considered, as their number is quite small (less than $10\%$ in the CosmoDC catalog) and thus their density profile may be significantly biased due to numerical effects and cosmic variance. Therefore, we do not involve these high redshift clusters as lenses in the later fitting procedures, but rather use the halo mass function to estimate their contribution to the total detection rate, which may introduce an extrapolation uncertainty by a factor of $\sim 1-2$. 
}

We cut out the field of each galaxy cluster into an area of $180^{\prime \prime} \times 180^{\prime \prime}$ toward a region dominated by the strong lensing effect. Each cluster is characterized by a dark matter halo described by the NFW profile, along with tens to hundreds of member galaxies described by the singular isothermal ellipsoid (SIE) models. Note that the mass distribution of the dark matter halo, the number and physical properties of member galaxies, such as velocity dispersion and ellipticity, and their relative position in the main halo are extracted directly from the catalog, produced by N-body simulation together with the semi-analytical modeling of galaxy formation and evolution in those halos. By numerical calculations for each cluster lensing system, we obtain discrete distribution of cross-section of the clusters from the CosmoDC2 mock catalog.

Figure~\ref{fig:clus} shows the critical/caustics curves (red/green curves) for an example lensing system extracted from the CosmoDC2 catalog, of which $M_\ell = 1.9\times 10^{15} M_{\odot}$ and $M_{\ast,\rm BCG}= 2\times 10^{11} M_{\odot}$. As seen from this figure, the lensing cross-section of the cluster is dominated by the central BCG and the main halo, while other member galaxies only provide small distortion/addition to the cross-section of the cluster lensing system.

According to our calculations, the cross-section for those mock clusters can be simply expressed as a function of the cluster mass and redshift, i.e., 
\begin{equation}
S_{\rm cr}(z_\ell,z_{\rm s},M_\ell)=f(M_\ell,z_\ell,z_{\rm s})R_{\rm E}^2(M_\ell,z_\ell)
\label{eq:Fit}
\end{equation}
where $f(M_\ell,z_{\rm s},z_\ell)$ is a monotonic function of $z_\ell$ within $(0.1,20]$, fixing other two parameters. The motivation of introducing this expression is that the values of $S_{\rm cr}$ for most galaxy clusters in the CosmoDC2 are on the order of $R_{\rm E}^2$ and the scale factor $f(M_\ell,z_{\rm s},z_\ell)$ is only strongly dependent on the physical properties of the DM halo and the source redshift. Note that the parameters of our catalog of galaxy cluster strong lensing events are limited to a small mass range $\sim (10^{14}-3\times 10^{15} M_\odot)$ and redshift range $z_{\ell} \in (0.2-1)$. To estimate the optical depth for cluster lensing across the cosmic time, we need to extrapolate the cross-section estimates to a larger mass range  (i.e., $\lesssim 10^{14} M_{\odot}$) and a larger redshift range  (i.e., $z_{\ell} \gtrsim 1.0$).

\end{itemize}

\begin{figure}
\centering
\includegraphics[width=1.0\columnwidth]{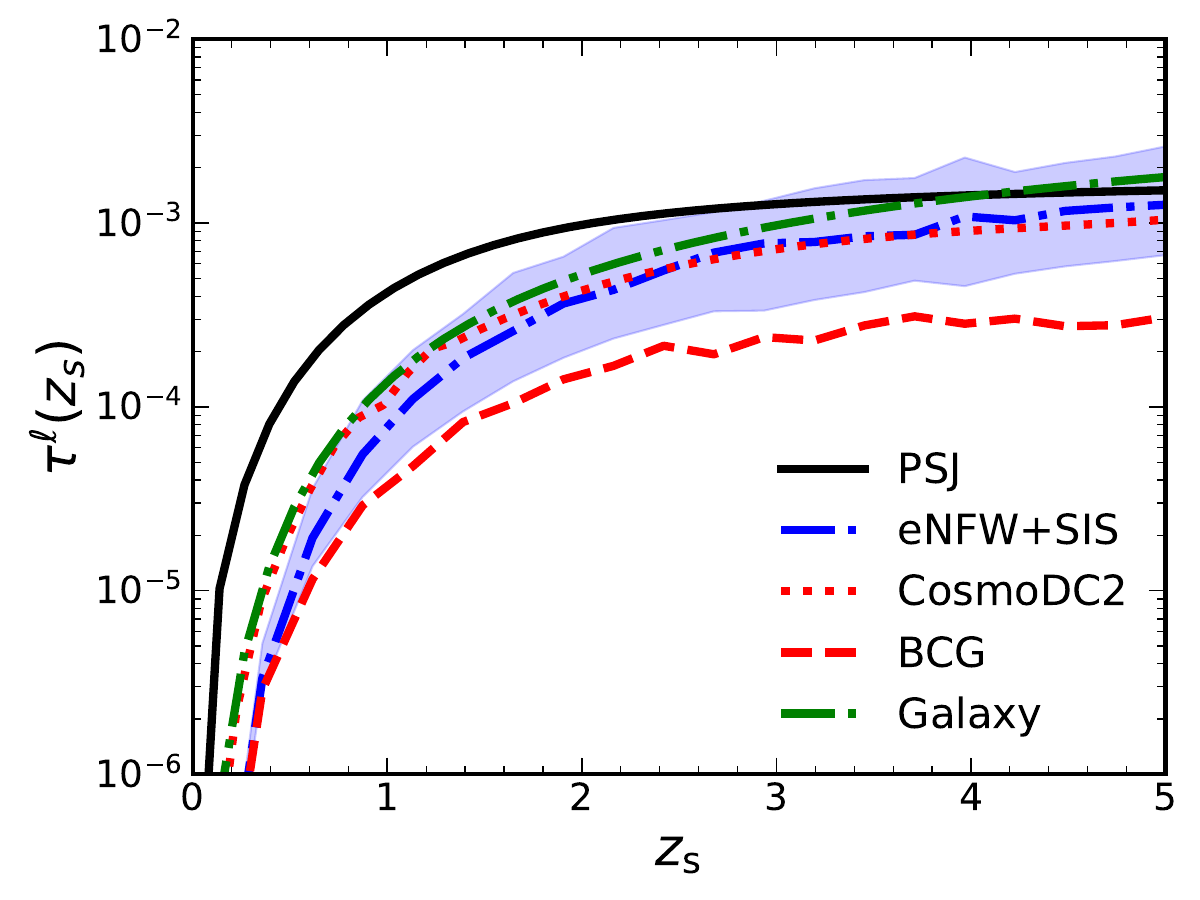}
\caption{The redshift evolution of the cosmic lensing optical depth. The black solid line  show the results adopting the PSJ profile. The upper and lower bound of the blue shaded region are the optical depth estimated by the eNFW+SIS profile adopting $c^{\rm o}_{\rm v}$ and $c^{\rm w}_{\rm v}$ relationship respectively, while the blue dotted-dashed line is the result adopting the average concentration of both, i.e.,  $c^{\rm a}_{\rm v}=(c^{\rm o}_{\rm v}+c^{\rm w}_{\rm v})/2$. The red dashed line shows the optical depth only considering the BCG in galaxy clusters. The red dotted line is the optical depth inferred from the CosmoDC2 catalog data  extrapolation equation~\ref{eq:Fit}. For a comparison, we also plot the optical depth for galaxy lensing from \citet{2023ApJ...953...36C} in green dot-dashed line. }
\label{fig:op}
\end{figure}

With the above lens models, we estimate the optical depth $\tau^{\ell}(z_{\rm s})$ by utilizing the Python package \textbf{hmf} \citep{2013A&C.....3...23M} and integrate Equation~\eqref{eq:tau} with the typical halo mass range of $M_{\ell}\in [10^{13},10^{15}] M_{\odot}$ for galaxy clusters. Figure~\ref{fig:op} shows $\tau^{\ell}(z_{\rm s})$ estimated from both the above different approaches, which are in general consistent with each other. On the one hand, adopting the PSJ profile gives the highest estimation of $\tau^{\ell}(z_{\rm s})$ compared with other results, especially at low redshift. On the other hand, one may see that $\tau^{\ell}(z_{\rm s})$ obtained by assuming the eNFW+SIS profile with the adoption of $c^{\rm w}_{\rm v}$ is about $\sim 4$ times smaller than that with $c^{\rm o}_{\rm v}$, indicating that the different choices of the $c_{\rm v}-M_\ell$ relationship  introduce a significant scatter to the estimation of $\tau^{\ell}(z_{\rm s})$. Nevertheless, we also estimate $\tau^{\ell}(z_{\rm s})$ assuming the eNFW+SIS model by adopting $c^{\rm a}_{\rm v}=(c^{\rm o}_{\rm v}+c^{\rm w}_{\rm v})/2$, an average concentration, and find the resulting $\tau^{\ell}(z_{\rm s})$ is generally consistent with that from the CosmoDC2 extrapolation (with deviation no more than a factor of $\sim 1.5$). This suggests that the simpler eNFW+SIS models for clusters may be as robust as the much more complicated models directly using the mock clusters in the CosmoDC2 catalog, but the former can be more flexible in analyzing the statistics of lensed GW events by clusters. We further calculate the optical depth due to BCGs in galaxy clusters without considering the dark matter halos and we find that the optical depth contributed by BCGs only is a factor of $\sim 4-5$ times smaller than the optical depth by considering both BCGs and the outer dark matter halos, which emphasize the importance of dark matter halo in the lensing effects of galaxy clusters. For comparison, we also show the optical depth for galaxy lensing estimated by \citet{2023ApJ...953...36C} in the Figure, which is somewhat larger than the value of cluster lensing. This can be explained by the larger number density, though smaller cross-section of the galaxies compared with galaxy clusters. Note that the member galaxies in the cluster contribute little to the total optical depth, less than $\sim 1\%$, though for a small number of certain clusters, the cross-section can vary by a factor of $1.5-2$. This indicates that the detection rate of lensed GW events can be well estimated by only considering the potential from the dark matter halo and BCG components of galaxy clusters but ignoring the contribution from member galaxies. 

Note that the cross-section $S_{\rm cr}$ of a galaxy cluster is directly related to its mass. By integrating Equation~\eqref{eq:tau} with different mass ranges, we find that the ratio of the optical depth obtained for clusters in the mass range of $10^{13}-10^{14}M_{\odot}$ to that in the range of $10^{14}-10^{15}M_{\odot}$ is about $1.5-4$, depending on the redshift of the source $z_{\rm s}$. Therefore, we conclude that the optical depth of galaxy cluster strong lensing is mainly contributed by low-mass galaxy clusters, which is mainly due to the dominance of the effect caused by the decline of cluster abundance with cluster mass. Note also that this estimation may be invalid for very low-mass dark matter halos. For example, at mass substantially lower than $10^{12}M_\odot$, the central galaxies may not be ellipticals but spirals with smaller concentration, and not each dark matter halo has a big central galaxy, thus contributing little to the lensing optical depth. 

\section{Results}
\label{sec:results}

\begin{table*}
\centering
\caption{
The detection rate of cluster-lensed sBBH mergers produced by both the EMBS and dynamical channels, adopting different cluster lensing settings, i.e., the pseudo-Jaffe density profile, the eNFW+SIS density profile ($c^{\rm a}_{\rm v}$, $c^{\rm o}_{\rm v}$ and $c^{\rm w}_{\rm v}$), and the extrapolation of lensing cross section obtained numerically for the clusters in the CosmoDC2 cataloge. 
For comparison, the detection rate of galaxy-lensed sBBH mergers and the total detectable sBBH merger event rate are listed in the last two columns.
Note here that the error of the detection rate is induced by the $90\%$ confidence uncertainty of the local merger rate density calibration. 
}
\begin{tabular}{lccccccc} \hline  \hline
$\dot{N}^{\ell}(\rm yr^{-1})$& PSJ   & CosmoDC2  & eNFW+SIS ($c^{\rm a}_{\rm v}$)   & eNFW+SIS ($c^{\rm o}_{\rm v}$) &  eNFW+SIS ($c^{\rm w}_{\rm v}$) & Galaxy & $\dot{N}_{\rm total}$ \\ \hline \hline
EMBS & $26_{-4.0}^{+58}$  & $13_{-2.0}^{+28}$ & $12_{-2.0}^{+27}$ & $23_{-4.0}^{+51}$  & $6_{-1.0}^{+14}$ & $14^{+30}_{-2.0}$ & $2.8_{-0.4}^{+5.9}\times 10^4$\\
Dynamical & $35_{-25}^{+10}$ &  $19_{-13}^{+5.0}$ & $19_{-14}^{+5.0}$ & $36_{-26}^{+10}$ & $9_{-7.0}^{+3.0}$ & $23_{-16}^{+7.0}$ & $3.2_{-2.3}^{+0.9}\times 10^4$\\ \hline \hline
\end{tabular}
\label{tab:1}
\end{table*}

\begin{figure}
\centering
\includegraphics[width=1.0\columnwidth]{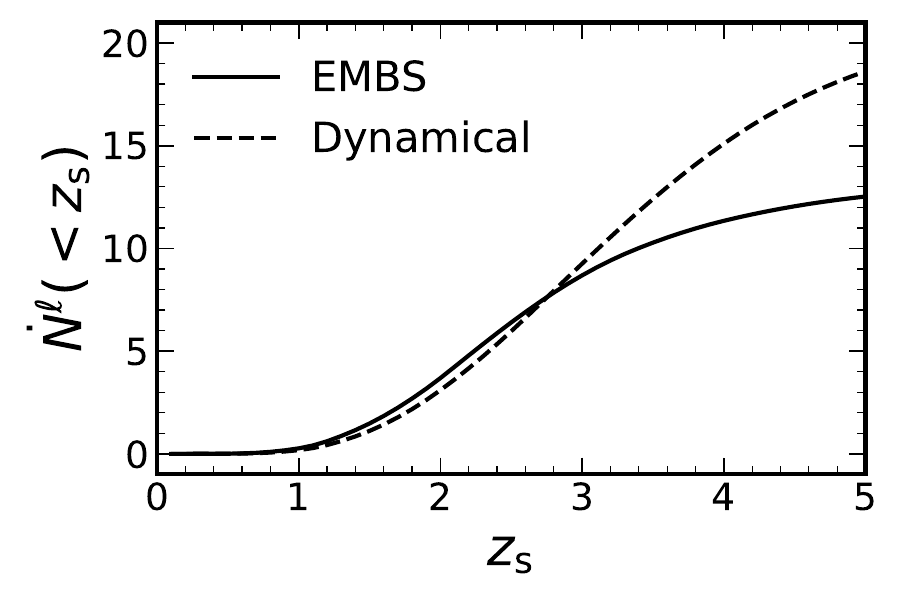}
\includegraphics[width=1.0\columnwidth]{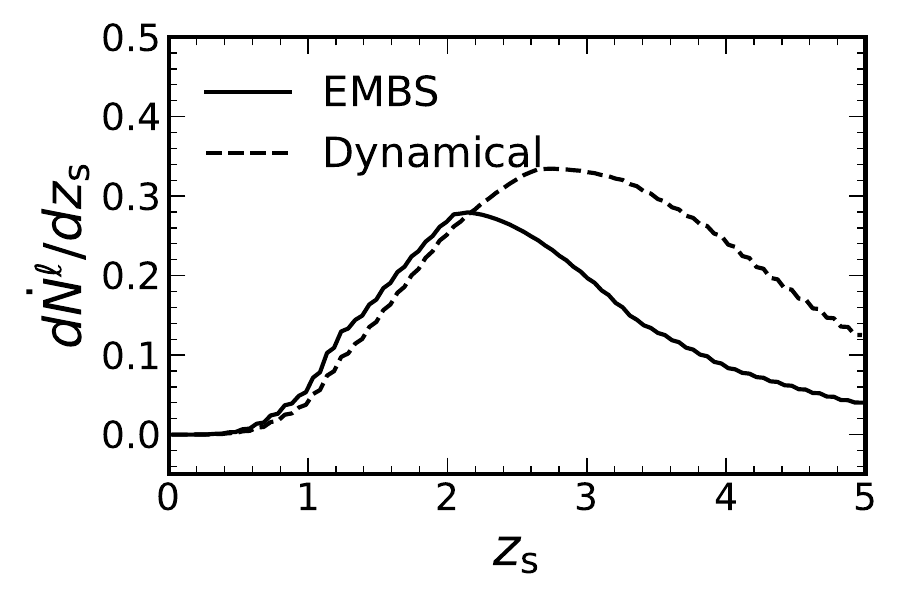}
\caption{The accumulative (top panel) and differential (bottom panel) source redshift distribution, i.e.,  $\dot{N}^{l}(<z_{\rm s})$ and $d\dot{N}^{l}/dz_{\rm s}$   of cluster lensed sBBH mergers produced by EMBS (solid) and dynamical (dashed) channel.
}
\label{fig:channel}
\end{figure}

By integrating over Equation~\eqref{eq:Nl}, we obtain the final estimation on the detection rate of sBBH mergers lensed by galaxy clusters. The results are shown in Table~\ref{tab:1}.  As for the PSJ model, the detection rate is relatively higher than the others, ${\sim 26_{-4.0}^{+6.0}/35_{-25}^{+10}}$\,yr$^{-1}$ for EMBS and dynamical channel respectively.  In this model, we did not consider detailed evolution of baryonic and DM density distribution of galaxy clusters, but given a constant core $(s,a)=(0.2,2) \xi_{0,\rm PSJ} $ by the typical scale of the modeling in real cluster lensing observations \citep{2015MNRAS.452.1437J,lotz17,annu17,2018arXiv180600698Q,2024MNRAS.531.1179X}, like HST-frontier field. Moreover, assuming the cross-section $S_{\rm cr}$ to be proportional to $M_{\ell}$ will overestimate the detection rate by a factor of few.  Therefore, we treat the estimation adopting the PSJ model as an upper limit of $\dot{N}^{\ell}$. As for the eNFW+SIS model, we note that different choices of concentration $c_{\rm v}$ will add $2-4$ times variance on the final detection rate. If adopting the $c^{\rm a}_{\rm v}$,  the detection rate $\dot{N}^{\ell}$ is about {${\sim12_{-2.0}^{+27}/19_{-14}^{+5.0}}$\,yr$^{-1}$}, which is about half of the value in the PSJ case, but very similar with  the results estimated based on the CosmoDC2 mock catalog, i.e., $\dot{N}^{\ell}\sim{ 13_{-2.0}^{+28}/19_{-13}^{+5.0}}$\,yr$^{-1}$. This indicates that adopting the  eNFW+SIS model is adequate for the rough estimation of the  detection rate of galaxy cluster lensed GW sources.   Here we have to noticed that though the CosmoDC2 catalog is dependent on the sub-grid physics of simulation and the pre-determined Halo Occupation function (HOF), we view the results as the most reliable estimation so far for the  detailed modeling of the structures of the galaxy clusters. 

As a comparison, we also list the detection rate of sBBH mergers strongly lensed by intervening galaxies, which is almost equivalent to the results of CosmoDC2, i.e., $\sim {14^{+30}_{-2.0}/23^{+7.0}_{-16}}$\,yr$^{-1}$ for EMBS and dynamical channel respectively. These results are in the same order of magnitude as the detection rate of cluster lensing, which can be partly explained by the larger cross-section of the galaxy clusters but rather smaller number density. These two factors lead to similar optical depth of galaxy and galaxy cluster lensing shown in Figure~\ref{fig:op} and thus a similar detection rate. Therefore, similar to  \citet{2023MNRAS.520..702S}, we conclude that in the next-generation GW detection, the galaxy and galaxy clusters may play equally important roles in the strong lensing of GW signals. 

In addition, we note that the expected detection rate can be different by a factor of $\sim 1.5$ for different formation channels. We also show the accumulated and differentiated source redshift distribution of the detectable lensed sBBH mergers produced by both the EMBS and dynamical channel in Figure~\ref{fig:channel}. As seen from the figure, most of the cluster lensed sBBH mergers are expected to be found at redshift $z_{\rm s}\sim 2$ for the EMBS channel, while $z_{\rm s}\sim 3$ for the dynamical channel. Similar to the discussion in \cite{2023ApJ...953...36C}, this discrepancy between the two formation channels can be explained by that the predicted intrinsic sBBH merger rate densities are different, due to different estimations of the local merger rate density and their evolution with redshift (see Fig.~\ref{fig:rbb}). Overall, the sBBH mergers lensed by galaxy clusters are more likely to be detected at higher redshift, rather than lower redshift. 

\section{Conclusions and Discussions}
\label{sec:con}

In this paper, we estimate the detection rate of strongly galaxy cluster-lensed sBBH mergers by the third-generation GW detectors. We adopt detailed modeling of the galaxy cluster lenses in the CosmoDC2 catalog and/or approximations of the PSJ profile or an eccentric NFW dark matter halo plus a bright central galaxy with SIS profile to calculate the lensing cross-section. We find that the involvement of the dark matter halo in the cluster can enhance the lensing cross-section by a factor up to $10$ (depending on the mass of the dark matter halo) compared with only considering the contribution of the central BCG. We consider both the EMBS channel and the dynamical channel for the formation of sBBH mergers and estimate the merger density rate by assuming that all sBBH mergers are either produced by the EMBS channel or the dynamical channel. Then we combine the halo mass function, lensing cross-sections, and the merger rate density to further estimate the lensing optical depth and the detection rate of lensed sBBHS for each lens model. Our main conclusions are summarized as follows.

\begin{itemize}
\item The lensing optical depth resulting from the eNFW+SIS model can vary by a factor of $\sim 2-4$ depending on different choices of the relationship between concentration and halo mass, but it is more or less consistent with the value estimated by using the more realistic CosmoDC2 galaxy clusters and the interpolation/extrapolation based on them. 
\item The detection rate of lensed sBBH mergers is expected to be in the range of $\sim 5-84$\,yr$^{-1}$ if all sBBH mergers are produced via the EMBS channel. The wide range of the estimated value reflects the uncertainties due to both the lens model and the error in the merger rate density (induced by the current observational uncertainty in the local merger rate density measurement).
\item Adopting the relatively more realistic galaxy clusters with central main and member galaxies in CosmoDC2 catalog for lenses, the resulting detection rate is about $\sim {13_{-2.0}^{+28}}$\,yr$^{-1}$ (with the superscript and subscript numbers indicating the errors due to the current observational uncertainty in the local merger rate density for calibration), which is close to the estimated detection rate of sBBH mergers lensed by galaxies.
\item The expected detection rate of lensed sBBHs for sBBHs produced by the dynamical channel is about a factor of $\sim 1.5$ larger than that by the EMBS channel. The redshift distribution of the lensed sBBHs from the dynamical channel peaks at a higher redshift $\sim 3$ compared with those from the EMBS channel (peaking at redshift $\sim 2$).
\end{itemize}

Note that there are many complexities one may need to take into account to make a more robust estimate. 
As for the lensing statistics in our analysis, the single-plane approximation is adopted to calculate the lensing cross-section of the galaxy clusters, but the effect of possible line of sight structures is ignored, which may introduce a small uncertainty to the detection rate \citep[e.g., ][]{2019ApJ...878..122L,2021MNRAS.504.2224L}. One may use multiple plane ray tracing method for clusters produced in large cosmological simulations, which may give more accurate estimates of the cluster lensing cross section. 
Furthermore, limited by the incompleteness of the CosmoDC2 catalog, the lensing cross-sections by clusters at lower mass and higher redshifts are obtained by extrapolation based on simple approximations and the optical depth of cluster lensing is estimated based on the fitting results. The choice of the fitting form may also introduce uncertainties to the estimation. One may use galaxy clusters produced by
(future) hydrodynamic simulations such as BAHAMAS \citep{2017MNRAS.465.2936M}, Millennium TNG \citep{2023MNRAS.524.2539P}, and FLAMINGO \citep{2023MNRAS.526.4978S} in larger cosmological volumes to do the lensing statistics, which may help to overcome the incompleteness problem. 
These simulations also incorporate more detailed sub-grid physics and may produce galaxy clusters to better fit the observations, and thus lead to more realistic estimates of lensing cross section and optical depth. 

\begin{figure}
\centering
\includegraphics[width=1.0\columnwidth]{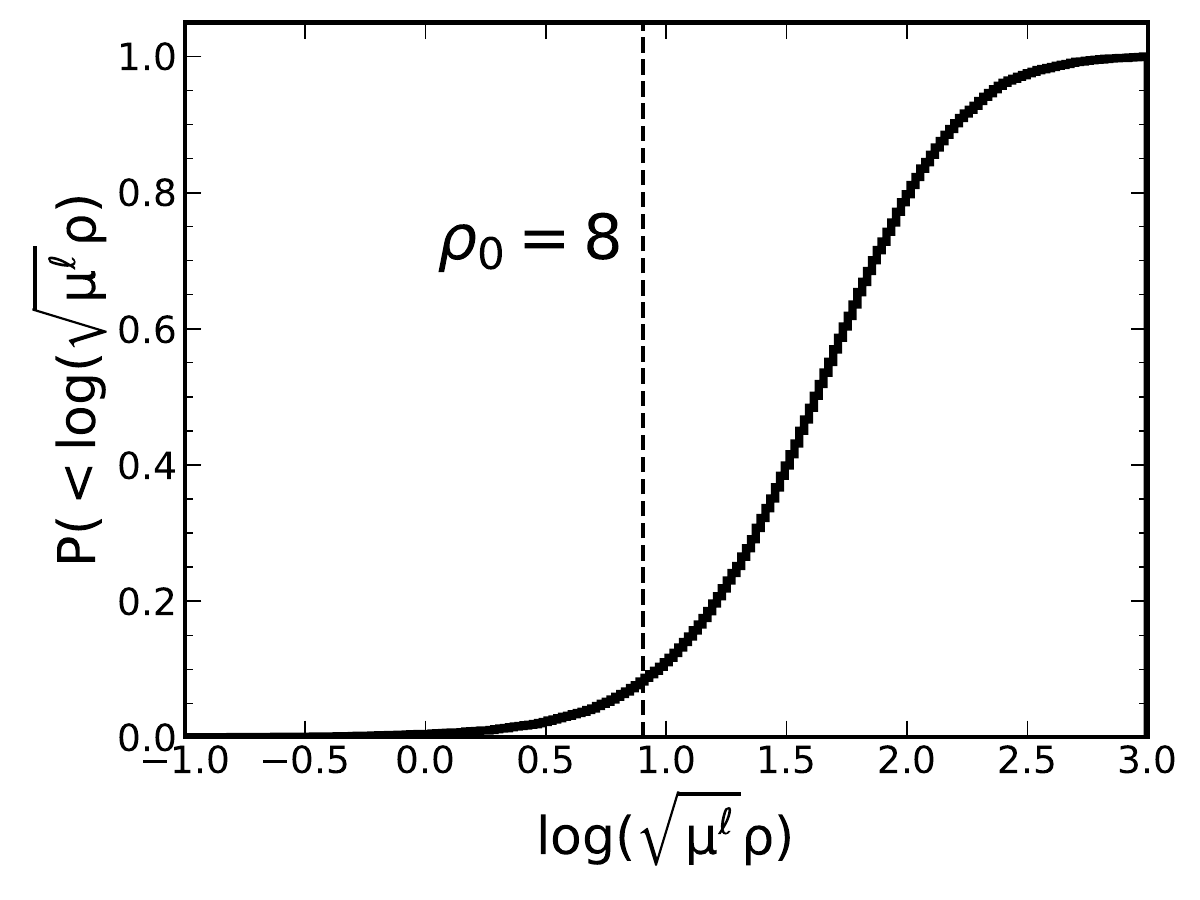}
\caption{The cumulative distribution of magnified SNR, i.e., $\sqrt{\mu^{\ell}}\rho$ of cluster lensed sBBH mergers produced via the EMBS channel, assuming the eNFW+SIS model. The dashed vertical line represent the typical threshold $\rho_0=8$ for GW detection. }
\label{fig:snr}
\end{figure}

As for the GW detection, the magnification bias was not considered, which may affect the detection rate estimation of lensed sBBHs. Due to the magnification by strong gravitational lensing, the S/N $\rho$ of distant or weak sources can be (de)-magnified by a factor of $\sqrt{\mu^\ell}$ with $\mu^{\ell}$ denoting the the magnification, which is significant for cluster lensing.  {Following the standard Monte Carlo procedure in \citet{2023ApJ...953...36C}, we estimate the SNR distribution of the lensed sBBH mergers produced via the EMBS channel, assuming the above eNFW+SIS lens model and the results are shown in Figure~\ref{fig:snr}. It can be seen that most lensed sBBH mergers, i.e., $\sim 92\%$ can be detected with CE, assuming the detection threshold to be $\rho_0=8$. As for those sBBH mergers produced via the dynamical channel, with relatively higher masses, the fraction is $\sim 95\%$. This indicates that the total detection rate of the lensed sBBH mergers may not change much if ignoring the magnification bias since almost all sBBH mergers within $z_{\rm s}\in [0,5]$ can be detectable by the third-generation GW detectors with high sensitivity. }

However, it becomes much more different for the cluster-lensing of binary neutron star (BNS) mergers, of which the S/N is relatively low, and thus the magnification bias becomes important. For example, for a cluster with halo mass $\sim 10^{14} M_{\odot}$ at redshift $z_{\ell}\sim 0.5$, the median magnification for sources at $z_{\rm s}\sim 1$ within the caustic is about $\mu^{\ell}\sim 3.82_{-1.00}^{+3.12}$ and $\mu^{\ell} \sim 2.41_{-0.33}^{+1.11}$ for the whole cluster and only the BCG respectively , with the subscript and superscript numbers indicating the $16\%-84\%$ errors. This result indicates that the lensed sBBH mergers possessing very high magnification are rare, which is more or less consistent with the prediction of the magnification distribution of LVK-O5 run in \citet{2023MNRAS.520..702S}. However, it is quite time-consuming to reconstruct the magnification factor map for each galaxy cluster. We defer the discussion on the impact of magnification bias in lensed BNS mergers for future work.  In addition, we only consider single formation mechanisms for the GW sBBH sources, i.e., either the dynamical or the EMBS channel. This may not be the case in the real universe, as both channels (and perhaps also other channels) could contribute to the sBBH merger population. Therefore, the total detection rate is dependent on the occupation fraction of each formation channel.

\section*{acknowledgement}
This work is partly supported by the National Natural Science Foundation of China (Grant No. 12273050 and 11991052), the Strategic Priority Program of the Chinese Academy of Sciences (Grant No. XDB0550300), and the National Key Program for Science and Technology Research and Development (Grant No. 2020YFC2201400 and 2016YFA0400704). HYS acknowledges the support from the Ministry of Science and Technology of China (grant Nos. 2020SKA0110100), NSFC of China under grant 11973070, Key Research Program of Frontier Sciences, CAS, Grant No. ZDBS-LY-7013, China Manned Space Project with NO. CMS-CSST-2021-A01, CMS-CSST-2021-A04. and  Program of Shanghai Academic/Technology Research Leader. 
NL acknowledge the support from the science research grants from the China Manned Space Project (No. CMS-CSST-2021-A01), the CAS Project for Young Scientists in Basic Research (No. YSBR-062) and the Ministry of Science and Technology of China (No. 2020SKA0110100).
XG acknowledges the fellowship of China National Postdoctoral Program for Innovative Talents (Grant No. BX20230104). 

\bibliographystyle{aasjournal}
\bibliography{ref.bib}
\appendix
\section{cross-section of Pseduo-Jaffe profile}
The density distribution of the Pseduo-Jaffe (PSJ) profile can be written as, 
\begin{equation}
    \rho(x)=\frac{\rho_s}{(x^2_s+x^2)(x^2_a+x^2)},
\end{equation}
where $x_s$ and $x_a$ is the dimensionless length of the core and transition radius. 
Here we have adopted the scale radius, $\xi_0$ to be:
\begin{equation}
    \xi^2_0=\frac{4(x_a+x_s)GM_{\ell}D_{\rm eff}}{x_a^2c^2},
\end{equation}
where $D_{\rm eff}=D_{\ell}D_{\rm s}/D_{{\ell}s}$ is the effective angular diameter distance of the lens system.
Then the dimensionless surface density (lensing convergence) can be calculated by projecting $\rho(x)$ to single plane, 
\begin{equation}
\kappa(x)=\frac{\kappa_s}{2}\left(\frac{1}{\sqrt{x_s^2+x^2}}-\frac{1}{\sqrt{x_a^2+x^2}}\right)
\end{equation}
where $\kappa_s={x_a^2}/{(x_a^2-x_s^2)}$ are the dimensionless properties of the PSJ profile.

The lensing potential $\psi(\boldsymbol{x})$ is directly related with the convergence $\kappa(\boldsymbol{x})$ by the Poisson equation, 
\begin{equation}
\nabla^2_{\boldsymbol{x}}\psi(\boldsymbol{x})=2\kappa(\boldsymbol{x})={\kappa_s}\left(\frac{1}{\sqrt{x_s^2+x^2}}-\frac{1}{\sqrt{x_a^2+x^2}}\right),
\end{equation}
where $x$ is the length of the position vector $\boldsymbol{x}$. Under the polar coordinate, the Possion equation can be solved and the lensing potential is   
\begin{equation}
\begin{aligned}
\psi(x)&=\kappa_s\Bigg[(\sqrt{x_s^2+x^2}-x_s)- (\sqrt{x_a^2+x^2}-x_a) \\
&-x_s\log{\Big(\frac{x_s+\sqrt{x_s^2+x^2}}{2x_s}\Big)}+x_a\log{\Big(\frac{x_a+\sqrt{x_a^2+a^2}}{2x_s}\Big)}\Bigg],
\end{aligned}
\end{equation}
assuming the central value $\psi(0)=0$ and $d\psi(x)/dx$ to be finite.

For the spherical symmetrical cases, the lens equation
can be reduced to 
\begin{equation}
    y=x-\frac{d\psi(x)}{dx},
\end{equation}
then the radius of the critical curve $x_{\rm crit}$ in the lens plane is defined by ${d^2\psi(x)}/{d^2x}=1$. Then the critical length in source plane (i.e., caustic) $y_{\rm crit}$ can be solved by the lensing equation.  For  demonstration purpose, in this paper we assume i.e., $(x_s,x_a)=(0.2,2)$ and find that $x_{\rm cr}=0.24$ and $y_{\rm cr}$ is about $\sim 0.17$.

\end{document}